\documentclass[10pt,twocolumn,letterpaper]{article}

\usepackage{cvpr}              

\usepackage{threeparttable}
\usepackage{longtable}
\usepackage{array}
\usepackage{pdflscape}

\definecolor{cvprblue}{rgb}{0.21,0.49,0.74}
\usepackage[pagebackref,breaklinks,colorlinks,allcolors=cvprblue]{hyperref}

\def\paperID{*****}
\def\confName{CVPR}
\def\confYear{2026}

\title{Drive-to-Music: Context-Aware Generative Audio for In-Vehicle Experiences}

\author{Cosmin Dragoiu \\
{\tt\small cosmin.dragoiu@mercedes-benz.com}
\and
Nooshin Nabizadeh \\
{\tt\small nooshin.nabizadeh@mercedes-benz.com}
\and
Mercedes-Benz Research \& Development North America
}

\begin{document}
\maketitle

\begin{abstract}
In-vehicle music can serve as an adaptive interface to enhance driver experience, attention, and well-being. We present Drive-to-Music, a context-aware system that generates music in real time from multimodal driving signals. Using dashcam imagery and vehicle telemetry, the system extracts scene semantics and driving context, maps them to high-level musical descriptors, and conditions generative audio models to produce contextually aligned soundtracks.

The architecture combines perception and generative components to translate visual and kinematic inputs into structured musical attributes and synthesize audio with low latency. It supports smooth transitions as driving conditions evolve, and to ensure robustness and deployment readiness, we incorporate constraint-based controls and safety checks across the generation pipeline.

Our results demonstrate the feasibility of real-time, context-aware music generation in automotive settings, providing a foundation for personalized and adaptive in-vehicle audio experiences.
\end{abstract}

\section{Introduction}

AI-driven music generation has become a prominent area of research, with broad applications across diverse domains~\cite{MitraZualkernan2025MusicGenReview}~\cite{YangLerch2020EvaluationMusicGen}~\cite{JainiKatikireddi2022MusicArtGenAI}. Systems that generate music from text, images, or video inputs are widely used across content creation industries such as digital production and stock music libraries~\cite{PlutPasquier2020GenerativeGames}. Despite these advances, the automotive industry remains an unexplored domain for AI music generation technology.

Music, long considered a passive companion during travel, has untapped potential as an adaptive and emotionally resonant medium that can enhance both focus and overall experience. SoundsRide~\cite{Kari2021SoundsRide}, an in-vehicle audio augmented reality system links environmental affordances to musical events using continuous prediction and real-time mixing. Mood-aware music recommendation for drivers~\cite{Cano2020MoodOnCar} infers driver mood and selects music intended to improve comfort and driving behavior. Studies on music effects on driving~\cite{Karageorghis2022InfluenceMusicDriving} show that music influences driver's cognition and style.

Despite these advances, most systems rely on selecting, reordering, or remixing existing music, and many evaluations are simulator-based prototypes rather than deployments in real vehicles. In this paper we introduce Drive-to-Music, a novel AI-driven system that dynamically generates personalized music based on real-time vehicle contextual data. Our system leverages dashcam imagery and vehicle telemetry to synthesize custom music tracks. By translating multimodal input into high-level musical descriptors, Drive-to-Music uses generative AI models to create audio output that is synchronized with the driving context.

In this paper, we outline the system architecture, multimodal input preprocessing and model selection. Our work contributes to the emerging field of context-aware generative music and sets the stage for future in-vehicle AI companions that can enhance driver well-being through personalized, intelligent soundscapes.

\section{System Architecture}

\begin{figure*}[htbp]
  \centering
  \includegraphics[width=0.8\linewidth]{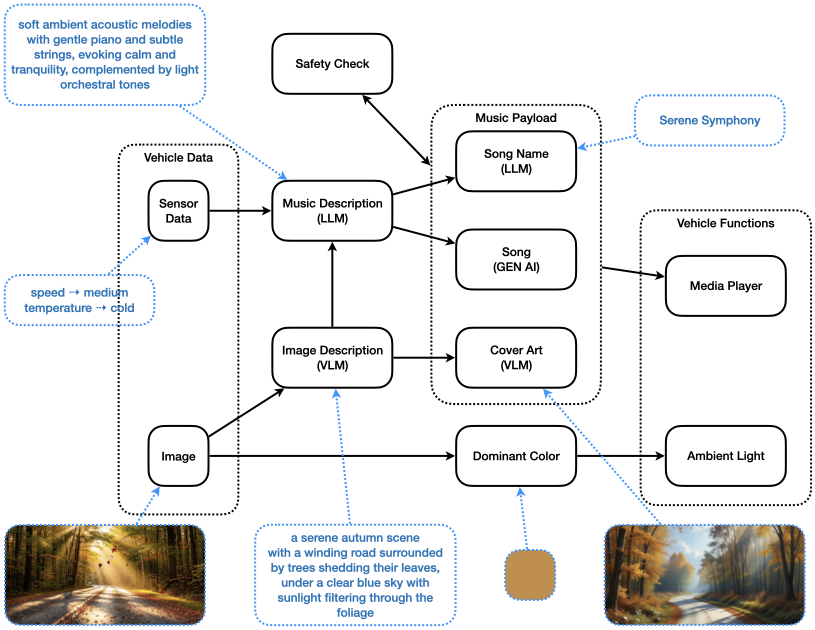}
  \caption{The Drive-to-Music system architecture showcasing the main components and the data flow.}
  \label{fig:d2m-architecture}
\end{figure*}

The Drive-to-Music architecture, shown in \autoref{fig:d2m-architecture}, consists of several stages that process multimodal vehicle data to produce contextually relevant music alongside complementary media elements such as cover art and song name. The main input to the system is a front-facing camera image that captures the driving scene. Additional vehicle sensor data, such as vehicle speed and outside temperature is used to further enhance the music generation process.

For a true in-vehicle experience, all generative AI models run in the cloud and a dedicated in-vehicle service is responsible for gathering the relevant data from the vehicle, sending it to the cloud, and receiving the generated music and media elements back to the vehicle for playback. For such a setup, the only requirement is an internet connection that enables periodic communications between the vehicle and the cloud.

\textbf{Image Description Stage}: To obtain a scene-level understanding of the driving context, we use a vision-language model (VLM) to generate an image description from the camera input. The prompt (see \autoref{tab:model_prompts}) is designed to extract information like surroundings, visible landmarks, vegetation, sky conditions, and weather. This description provides a compact semantic representation of the visual environment and serves as the basis for downstream generation. 

\textbf{Music Description Stage}: The scene description is passed to a large language model (LLM), which combines it with additional vehicle sensor data (see \autoref{tab:model_prompts}) and generates a set of music-oriented descriptors that best match the observed environment. The music descriptors capture attributes such as mood, atmosphere, and style, and are used to condition a music-generation model.

\textbf{Music Payload Generation Stage}: In parallel with music generation, the system produces additional media elements to provide a more complete playback experience. A cover-art image is generated from the previously extracted scene description using an image-generation model. The objective is to preserve the overall mood and semantic content of the original scene while allowing for a more stylized and flexible visual output (see \autoref{tab:model_prompts}). A separate LLM call is used to generate a short music title based on the music description.

\textbf{Dominant Color Extraction Stage}: To further enhance the in-vehicle experience, the system also extract the dominant color of the captured scene. This component uses a histogram-based approach. Muted sky regions and low-information gray areas, such as roads and building, are masked. The remaining pixels are mapped to a supported in-vehicle color palette, and the resulting dominant color is extracted. This color is used to dynamically adjust the ambient lighting in the vehicle, creating a more immersive and harmonious environment that complements the generated music and visual elements.

\textbf{Safety Check Stage}: All AI-generated artifacts are validated against specific safety and quality requirements. The generated music is checked to ensure that it is natural and free from abrupt jumps or excessive noise. The validation stack includes music quality scoring, loudness checks, true-peak checks, sudden-jump detection, and noise-floor analysis. The cover art is screened to ensure that it does not contain offensive visual content. Similarly, the generated music title is filtered to remove vulgar or otherwise disallowed language. In case a component fails the safety checks, the system regenerates it until it meets the requirements.

\section{Model Selection and Evaluation}

Across all components, model and algorithm selection is guided by key system-level constraints, including minimizing end-to-end latency, maintaining acceptable output quality, and reducing overall resource usage. The estimation of different metrics is based on evaluating on average about ten samples per model.

To select the best VLM for our system, we conducted a benchmarking study of several candidate models, including \textit{LiquidAI LFM 2.5 VL 1.6B}~\cite{liquidai2025lfm2}, \textit{Qwen 3 VL 2B Instruct}~\cite{qwen3technicalreport}, and \textit{Gemini 2.5 Flash Lite}~\cite{geminitechnicalreport}. We evaluated these models based on their inference latency, output quality, including the CLIP score~\cite{hessel2022clipscorereferencefreeevaluationmetric}, and resource usage. The latency is measured as the average time to process a single request. The results are summarized in \autoref{tab:vlm_benchmark}.

All models provide comparable output quality, with CLIP scores in the 26–27 range. However, there are significant differences in inference latency. \textit{Gemini 2.5 Flash Lite} achieves the lowest latency at approximately 0.6 seconds per request, while \textit{LiquidAI LFM2.5 VL 1.6B} and \textit{Qwen3 VL 2B Instruct} require around 1.2 and 3.9 seconds, respectively. Based on these results, we selected \textit{LiquidAI LFM2.5 VL 1.6B} as the scene-description component of our pipeline.

\begin{table}[htbp]
  \scriptsize
  \centering
  \begin{threeparttable}
  \begin{tabular}{lrcc}
    \toprule
    \textbf{Model} & \textbf{Capacity} & \textbf{Latency} & \textbf{CLIP Score} \\
    \midrule
    \textbf{LiquidAI LFM 2.5 VL 1.6B} & 1.6B & 1.2s \tnote{*} & 26.43 \\
    Qwen 3 VL 2B Instruct & 2B & 3.9s \tnote{*} & 26.59 \\
    Gemini 2.5 Flash Lite & 150B & 0.6s \tnote{$\dagger$} & 27.25 \\
    \bottomrule
  \end{tabular}
  \begin{tablenotes}
    \footnotesize
    \item[*] Local hosting on NVIDIA T4 (16GB)
    \item[$\dagger$] API call
  \end{tablenotes}
  \caption{Benchmark results for image understanding VLM candidates. The CLIP score is used to evaluate the semantic alignment between the generated description and the input image.}
  \label{tab:vlm_benchmark}
  \end{threeparttable}
\end{table}

For the music description generation stage, we benchmarked the following LLM candidates: \textit{LiquidAI LFM 2.5 1.2B Instruct}~\cite{liquidai2025lfm2}, \textit{Qwen 3.5 2B}~\cite{qwen3technicalreport}, and \textit{Gemini 2.5 Flash Lite}~\cite{geminitechnicalreport}. All models provide high output quality, with \textit{LiquidAI LFM 2.5 1.2B Instruct} and \textit{Gemini 2.5 Flash Lite} achieving a perfect score of 5/5. In terms of latency, \textit{LiquidAI LFM 2.5 1.2B Instruct} is the fastest at approximately 0.7 seconds per request, followed by \textit{Gemini 2.5 Flash Lite} at 1.2 seconds, and \textit{Qwen 3.5 2B} at 2.4 seconds. The complete benchmark results are summarized in \autoref{tab:llm_benchmark}.

Based on these results, the \textit{LiquidAI LFM 2.5 1.2B Instruct} model stands out and therefore was selected as the music description generation component of our pipeline. The same model is also used for the song name generation, as it provides a good balance of latency and output quality for both tasks.

\begin{table}[htbp]
  \scriptsize
  \centering
  \begin{threeparttable}
  \begin{tabular}{lrcc}
    \toprule
    \textbf{Model} & \textbf{Capacity} & \textbf{Latency} & \textbf{Quality} \\
    \midrule
    \textbf{LiquidAI LFM 2.5 1.2B Instruct} & 1.2B & 0.7s \tnote{*} & 5/5 \\
    Qwen 3.5 2B & 2B & 2.4s \tnote{*} & 4/5 \\
    Gemini 2.5 Flash Lite & 150B & 1.2s \tnote{$\dagger$} & 5/5 \\
    \bottomrule
  \end{tabular}
  \begin{tablenotes}
    \footnotesize
    \item[*] Local hosting on NVIDIA T4 (16GB)
    \item[$\dagger$] API call
  \end{tablenotes}
  \caption{Benchmark results for music description LLM candidates. The output quality is evaluated based on stability and relevance based on a set of curated test cases on a 5-point scale.}
  \label{tab:llm_benchmark}
  \end{threeparttable}
\end{table}

The music generation stage is at the core of the Drive-to-Music system and is the most computationally intensive component. The selection of the music generation model relies on a combination of factors. We prioritize models that can produce instrumental music with no vocals, as vocals can be a distracting factor for drivers. We also require that models be able to generate songs of 90 seconds or longer to minimize the frequency of song transition and to provide a more immersive experience.

From all available models, we selected \textit{ElevenLabs}~\cite{ElevenLabs}, \textit{Mubert}~\cite{Mubert} and \textit{Stable Audio 2.5}~\cite{StableAudio} as the most promising candidates for further evaluation. We benchmarked the selected models against latency and several audio quality criteria, which are listed in \autoref{tab:music_benchmark}. 

The MuQ-MuLan score~\cite{zhu2025muqselfsupervisedmusicrepresentation} measures the prompt–audio semantic alignment. It relies on a joint music–text embedding model trained contrastively for music-text similarity. The Technical Audio Quality Score (TAQS) is a custom 0–100 score that summarizes the engineering quality of an audio track based on loudness, true peak, clipping, and sudden loudness jumps. A higher score means the track is technically cleaner, safer for playback, and more production-ready.

Based on these criteria, we selected \textit{Stable Audio 2.5} as the music generation model for our system.

\begin{table}[htbp]
  \scriptsize
  \centering
  \begin{threeparttable}
  \begin{tabular}{lrcc}
    \toprule
    \textbf{Model} & \textbf{Latency \tnote{*}} & \textbf{MuQ-MuLan Score} & \textbf{TAQS Score} \\
    \midrule
    ElevenLabs & 15.7s & 0.09 & 72.22 \\
    Mubert & 7.4s & 0.29 & 91.26 \\
    \textbf{Stable Audio 2.5} & 6.2s & 0.36 & 78.03 \\
    \bottomrule
  \end{tabular}
  \begin{tablenotes}
    \footnotesize
    \item[*] API call
  \end{tablenotes}
  \caption{Benchmark results for music generation model candidates. The MuQ-MuLan score measure the prompt–audio semantic alignment. A high score means the music matches the text semantically. The TAQS score summarizes the engineering quality of an audio track. A higher score means a higher quality.}
  \label{tab:music_benchmark}
  \end{threeparttable}
\end{table}

In addition to the audio track, the system generates a companion cover art image. The goal is to preserve the essence of the captured front-facing camera scene without reproducing the original frame. To meet these requirements, we use a dedicated VLM that is optimizied for image generation. We tested several models, including \textit{Stable Diffusion 3.5 Large Turbo}~\cite{esser2024scalingrectifiedflowtransformers}, \textit{Qwen Image}~\cite{wu2025qwenimagetechnicalreport} and \textit{Gemini 2.5 Flash Image}~\cite{geminitechnicalreport} (see \autoref{tab:cover_art_benchmark}).

Similar to previous stages, the models were evaluated based on their inference latency, output quality, including the GenEval score~\cite{kamath2025geneval2addressingbenchmark}, and resource usage. The GenEval score measures how reliably a text-to-image model follows prompt composition, like object presence, count, color, position, and attribute binding, using automatic detectors instead of human judgment. \textit{Gemini 2.5 Flash Image} and \textit{Qwen Image} achieve the highest GenEval scores, with 0.96 and 0.91, respectively, while \textit{Stable Diffusion 3.5 Large Turbo} achieves a score of 0.66.

In terms of latency, \textit{Stable Diffusion 3.5 Large Turbo} is the fastest model, with an average latency of 1.4 seconds per request when running locally, followed by \textit{Gemini 2.5 Flash Image} at 6.5 seconds, and \textit{Qwen Image} at more than 40 seconds. Based on these results, we selected \textit{Stable Diffusion 3.5 Large Turbo} as the cover art generation component.

\begin{table}[htbp]
  \scriptsize
  \centering
  \begin{threeparttable}
  \begin{tabular}{lrrc}
    \toprule
    \textbf{Model} & \textbf{Capacity} & \textbf{Latency} & \textbf{GenEval Score} \\
    \midrule
    \textbf{Stable Diffusion 3.5 Large Turbo} & 8B & 1.4s \tnote{*} & 0.66 \\
    Qwen Image & 20B & 40.4s \tnote{*} & 0.91 \\
    Gemini 2.5 Flash Image & 150B & 6.5s \tnote{$\dagger$} & 0.96 \\
    \bottomrule
  \end{tabular}
  \begin{tablenotes}
    \footnotesize
    \item[*] Local hosting on NVIDIA A100 (80GB)
    \item[$\dagger$] API call
  \end{tablenotes}
  \caption{Benchmark results for cover art VLM candidates. The GenEval score measures how reliably a text-to-image model follows the prompt composition.}
  \label{tab:cover_art_benchmark}
  \end{threeparttable}
\end{table}

\section{Conclusions}

In this paper, we introduced Drive-to-Music, a context-aware system that generates personalized music from multimodal driving signals. By combining vision-language models for scene understanding, language models for mapping context to musical descriptors, and generative audio models for synthesis, the system translates driving conditions into adaptive soundtracks in real time, with representative examples provided in \autoref{tab:example_outputs}.

Through systematic evaluation, we selected components that balance latency, quality, and resource constraints, demonstrating the feasibility of integrating generative AI into in-vehicle experiences. Limited user studies showed positive reception, with participants appreciating the contextual relevance and personalization of the generated music. Future work will extend the system with additional signals, such as traffic, navigation, and user preferences, to further enhance personalization and contextual relevance.

{
    \small
    \bibliographystyle{ieeenat_fullname}
    \bibliography{main}
}

\newpage

\appendix
\onecolumn

\begin{landscape}

\section{Examples of Generated Content}
\label{sec:examples}

\renewcommand{\arraystretch}{2}
\begin{longtable}{
  >{\raggedright\arraybackslash}m{3.5cm}
  >{\raggedright\arraybackslash}m{2.5cm}
  >{\raggedright\arraybackslash}m{5cm}
  >{\raggedright\arraybackslash}m{5cm}
  >{\raggedright\arraybackslash}m{2.5cm}
}
  \toprule
  \textbf{Camera Image} &
  \textbf{Vehicle Data} &
  \textbf{Image Description} &
  \textbf{Music Description} &
  \textbf{Cover Art} \\
  \midrule
  \includegraphics[width=\linewidth]{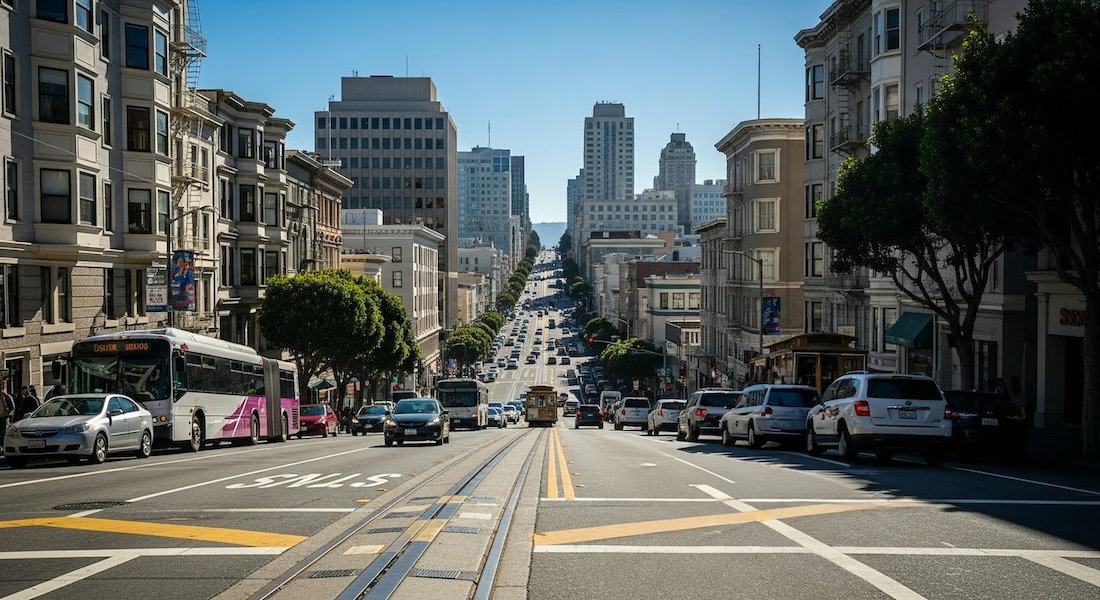} &
  low speed and warm outside temperature &
  a bustling urban street scene with a clear blue sky, tall buildings, and a tree-lined sidewalk, suggesting a sunny day in a city environment &
  upbeat indie folk with gentle rhythms, soft acoustic guitar, and subtle electronic undertones to match the lively yet relaxed atmosphere of a sunny urban street &
  \includegraphics[width=\linewidth]{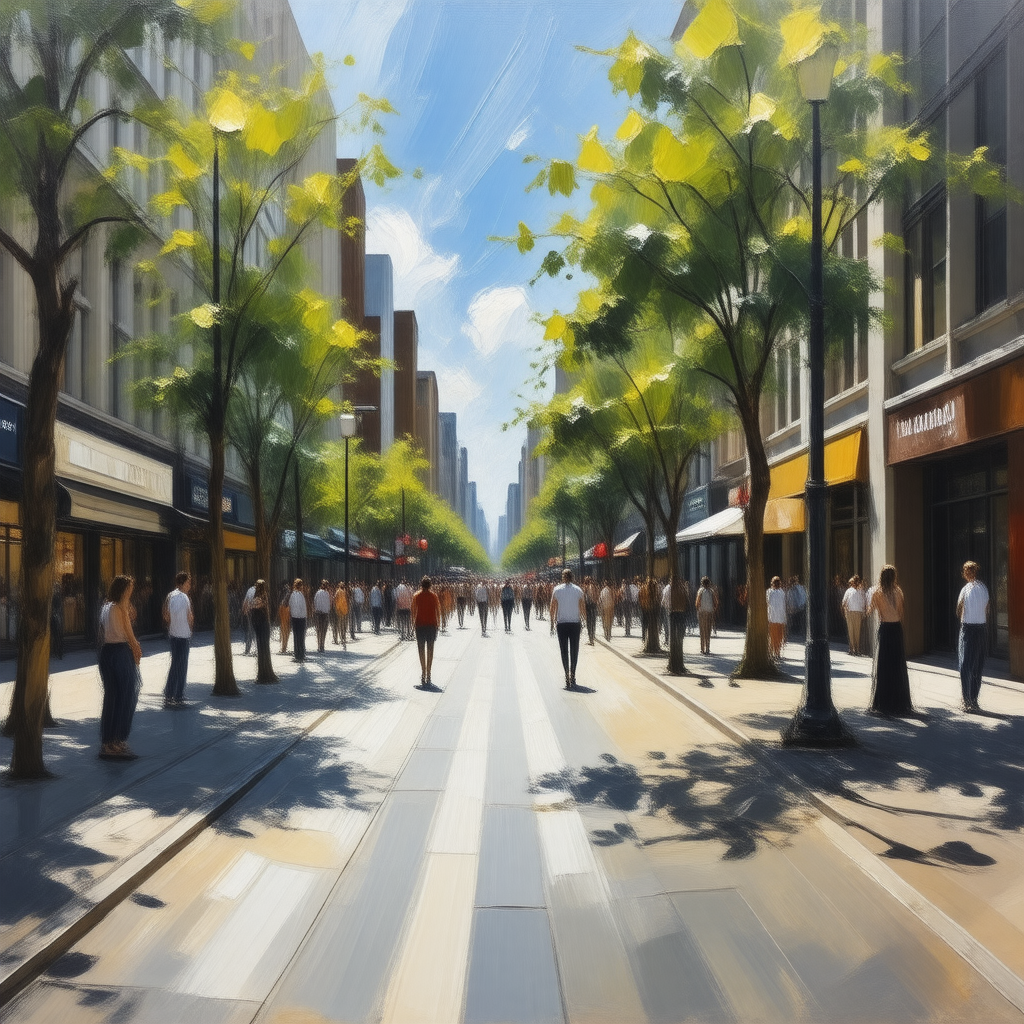} \\
  \includegraphics[width=\linewidth]{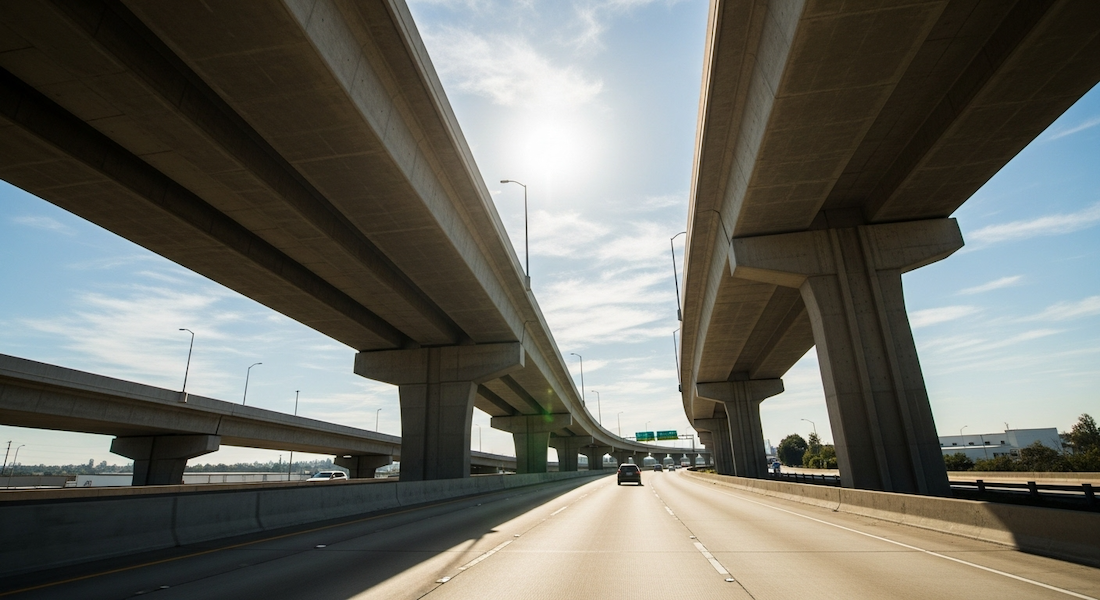} &
  high speed and hot outside temperature &
  a highway with multiple lanes, surrounded by concrete overpasses, under a clear blue sky with some clouds and sunlight shining through &
  high-energy electronic or synthwave tracks with driving rhythms, vibrant basslines, and uplifting melodies to match the speed and atmosphere &
  \includegraphics[width=\linewidth]{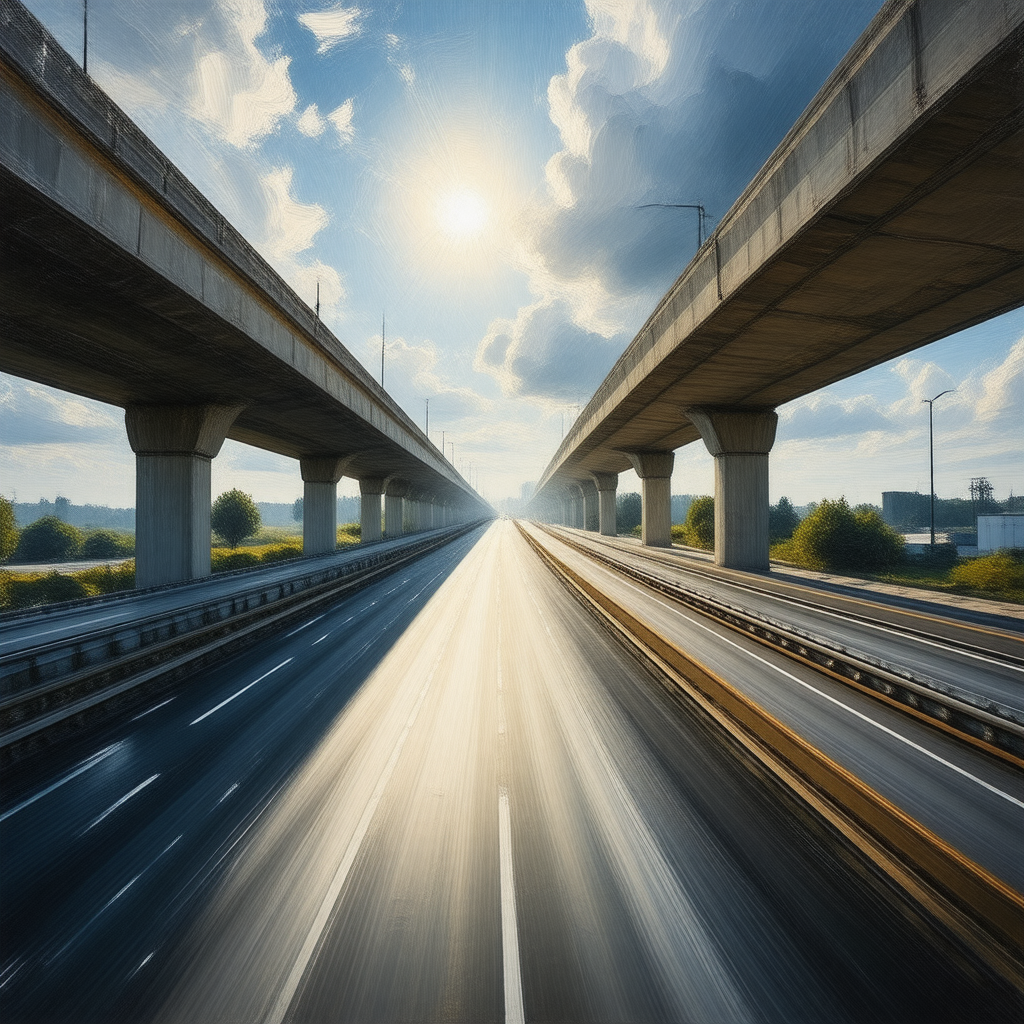} \\
  \includegraphics[width=\linewidth]{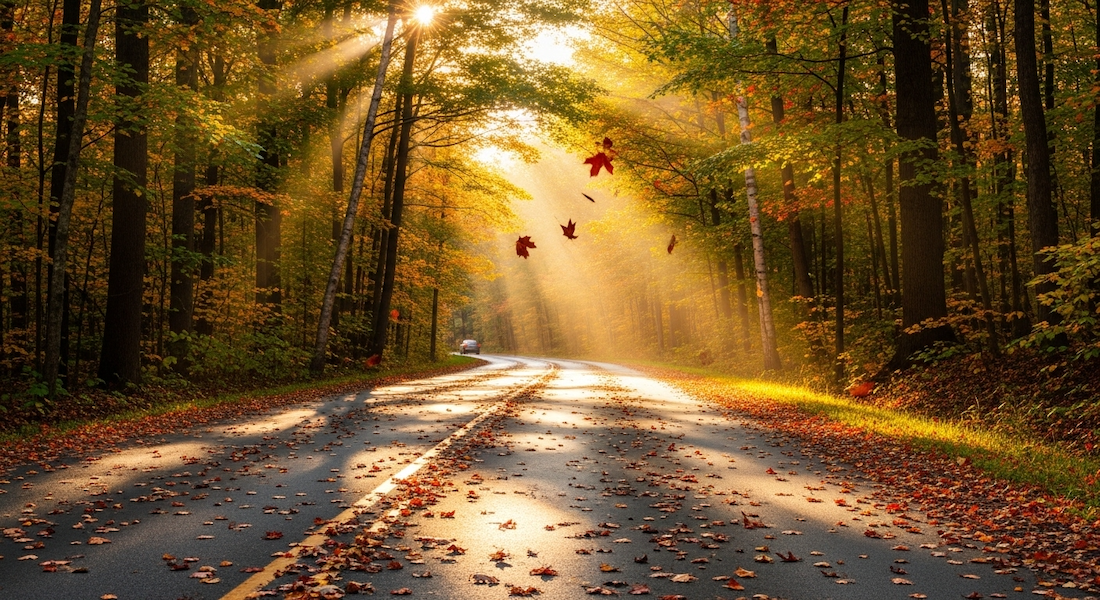} &
  medium speed and warm outside temperature &
  a serene autumn scene with a winding road surrounded by trees shedding their leaves, under a clear blue sky with sunlight filtering through the foliage &
  soft ambient acoustic melodies with gentle piano and subtle strings, evoking calm and tranquility, complemented by light orchestral tones &
  \includegraphics[width=\linewidth]{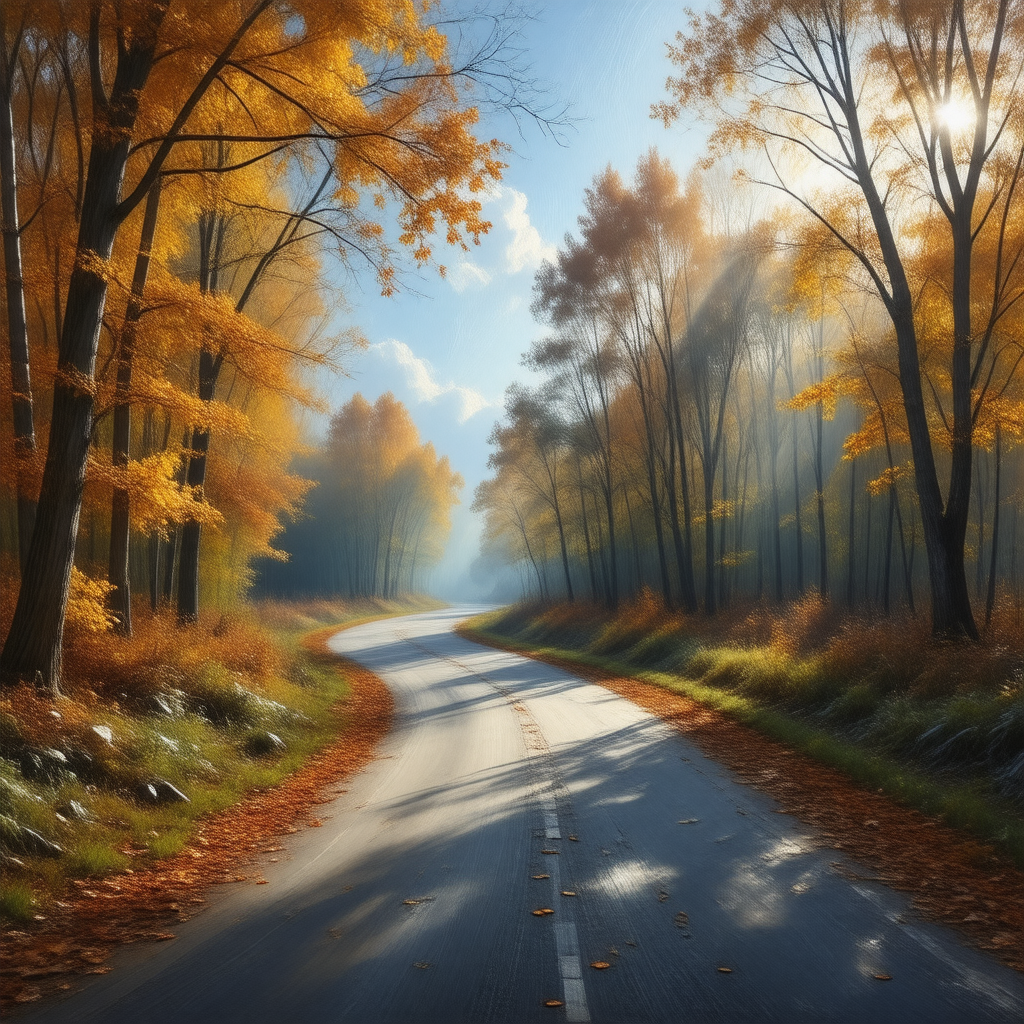} \\
  \includegraphics[width=\linewidth]{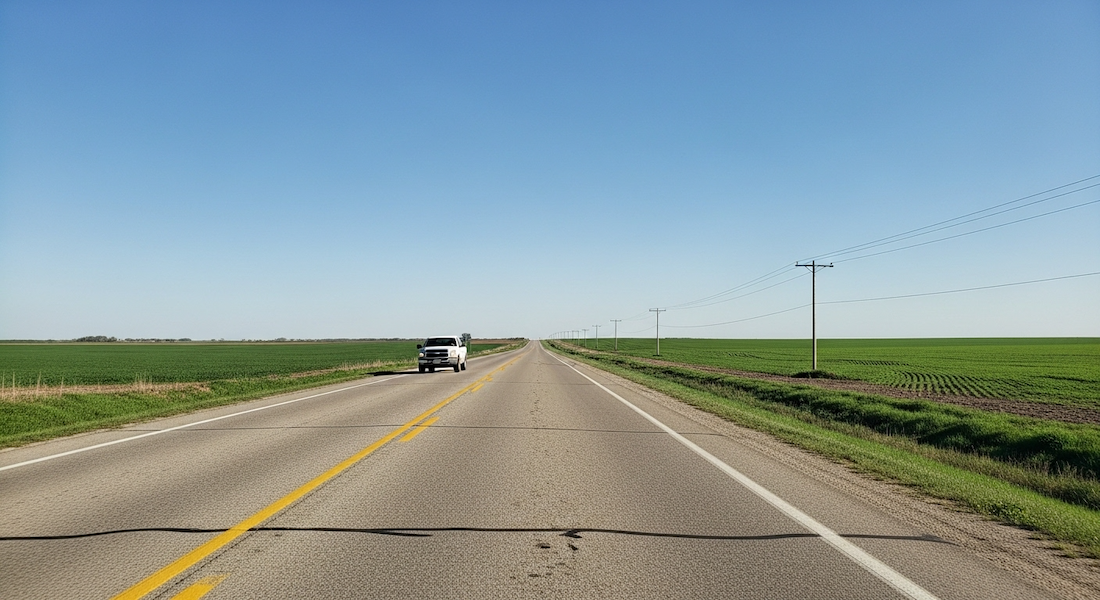} &
  high speed and warm outside temperature &
  a two-lane highway with a white truck traveling in the left lane, surrounded by green fields and power lines under a clear blue sky &
  high-energy electronic rock with driving rhythms, intense bass, and uplifting melodies to match the speed and atmosphere &
  \includegraphics[width=\linewidth]{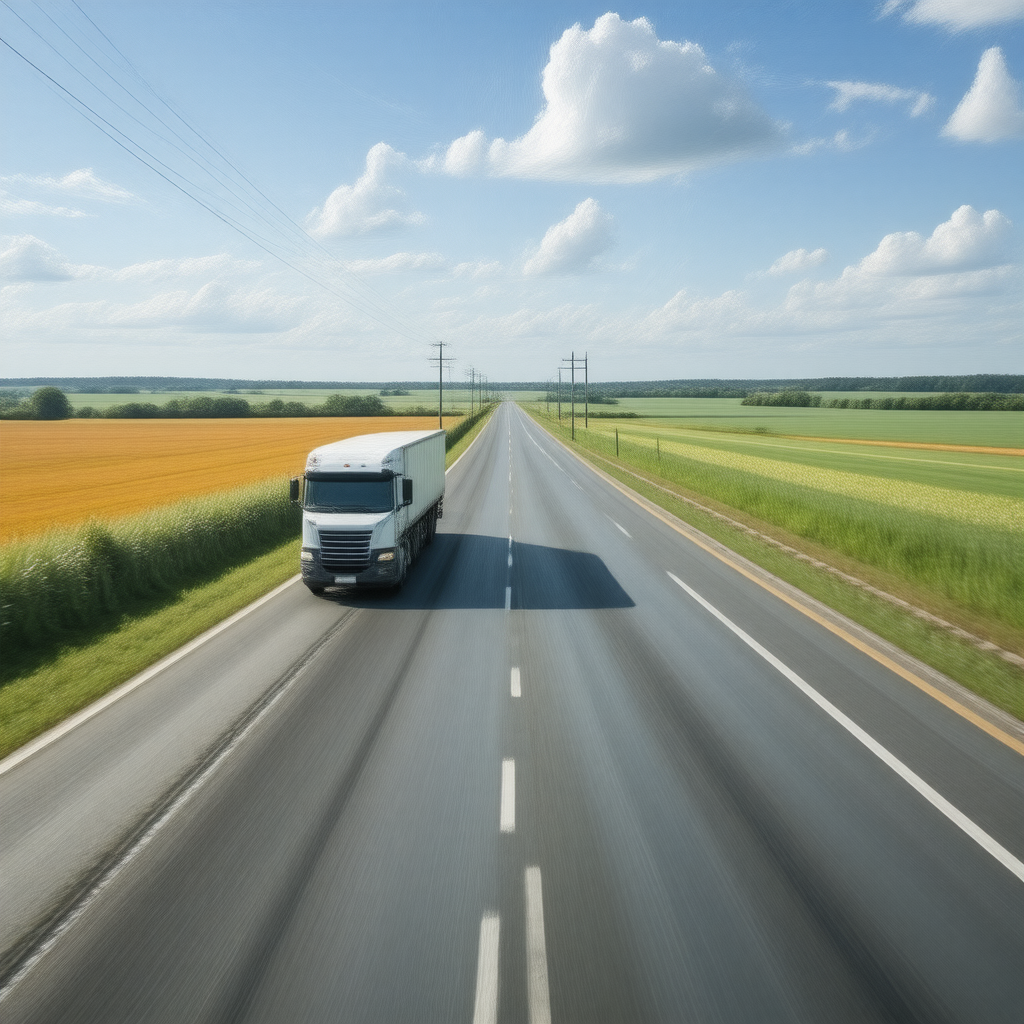} \\
  \includegraphics[width=\linewidth]{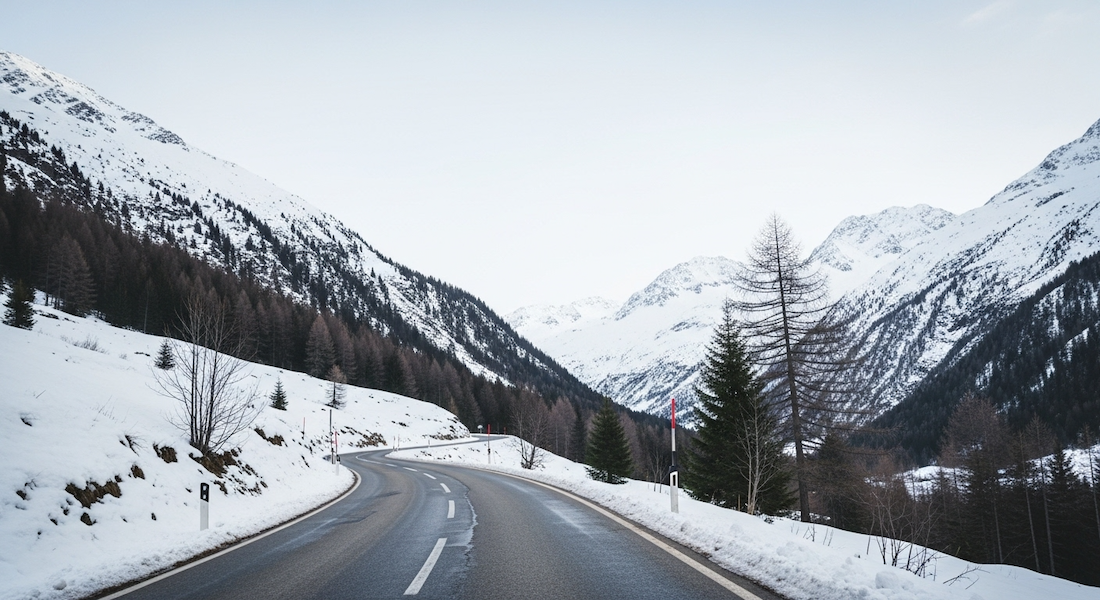} &
  medium speed and cold outside temperature &
  a snowy mountain road with a single lane and a guardrail, surrounded by snow-covered trees and mountains, under a clear blue sky &
  soft ambient acoustic melodies with gentle piano and subtle strings, evoking serene winter landscapes, paired with light electronic textures for a calm, reflective mood &
  \includegraphics[width=\linewidth]{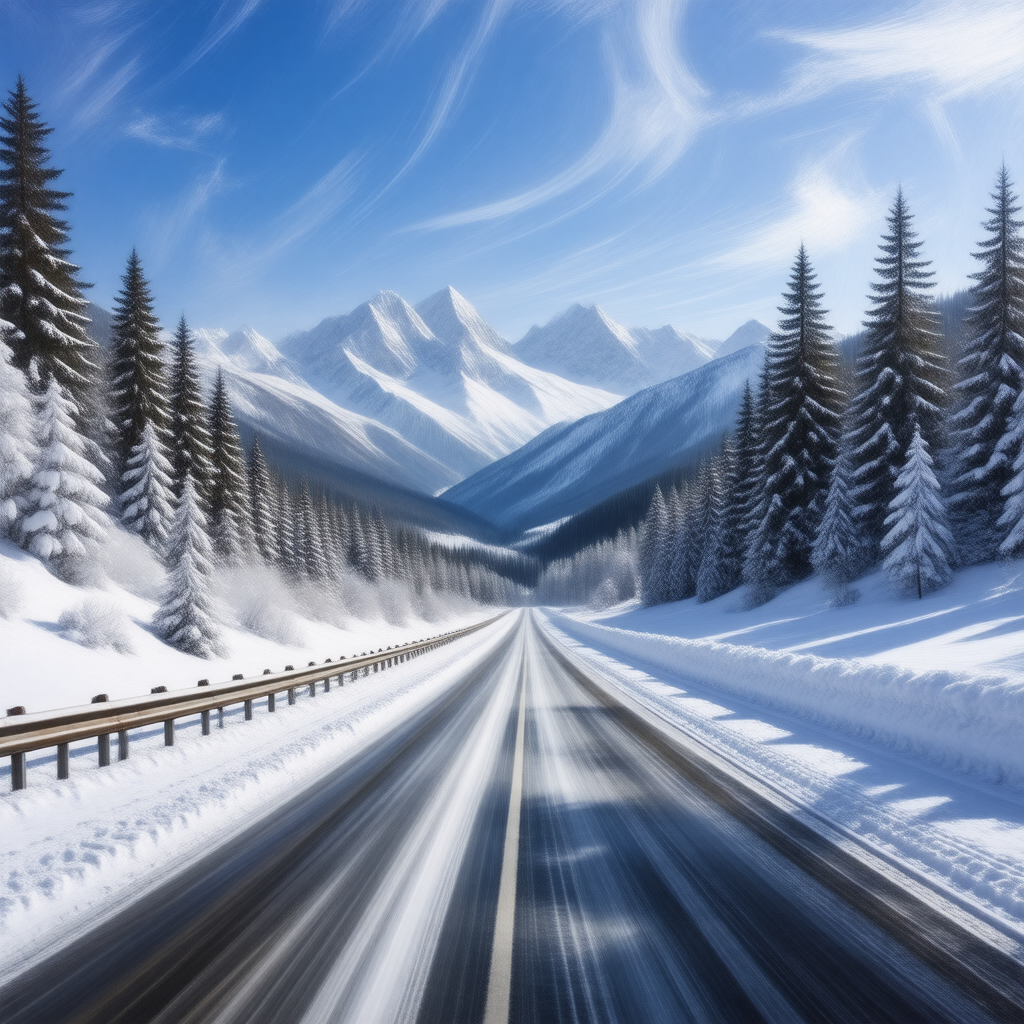} \\
  \bottomrule
  \caption{Examples of Drive-to-Music outputs showing scene descriptions mapped to musical characteristics and generated cover art images.}
  \label{tab:example_outputs} \\
\end{longtable}

\end{landscape}

\section{Model Prompts}
\label{sec:prompts}

\begin{table}[htbp]
  \centering
  \begin{tabular}{p{0.25\linewidth}p{0.65\linewidth}}
    \toprule
    \textbf{Model} & \textbf{User Prompt} \\
    \midrule
    Image Description VLM & describe the image by focusing on surroundings, landmarks, vegetation, weather; ignore people and offensive content in the image; limit the response to one sentence \\
    \addlinespace[0.5cm]
    Cover Art VLM & generate a colorful image depicting \textless{}image\_description\textgreater{}; do not generate people or offensive content; return an image with an aspect ratio of one \\
    \addlinespace[0.5cm]
    Music Description LLM & what kind of music would fit best \textless{}image\_description\textgreater{}?; take into account that the car drives at \textless{}speed\_level\textgreater{} speed and outside is \textless{}temperature\_level\textgreater{}; return only the music description, no other text \\
    \addlinespace[0.5cm]
    Song Name LLM & generate a short music title that captures the essence of \textless{}music\_description\textgreater{}; the title should be in Title Case style; return only the music title, no other text \\
    \addlinespace[0.5cm]
    Song Generation GenAI & generate \textless{}music\_description\textgreater{} \\
    \bottomrule
  \end{tabular}
  \caption{Prompts used for each model component as part of the Drive-to-Music pipeline.}
  \label{tab:model_prompts}
\end{table}

\end{document}